\documentclass[twocolumn,showpacs,amsmath,amssymb]{revtex4}
\usepackage{graphicx}
\begin{document}
\title{Polydisperse chains placed on a one-dimensional lattice}
\author{J\"urgen F. Stilck}
\email{jstilck@if.uff.br}
\author{Minos A. Neto}
\email{minos@if.uff.br}
\author{Wellington Gomes Dantas}
\email{wgd@if.uff.br}
\affiliation{Instituto de F\'{\i}sica\\
Universidade Federal Fluminense\\
Av. Litor\^anea s/n\\
24210-340 - Niter\'oi, RJ\\
Brazil}
\date{\today}

\begin{abstract}
Using a transfer matrix technique, we calculate the entropy of polydisperse
chains placed on an one-dimensional lattice, as a function of the density of
internal and endpoint monomers. The polydispersivity is determined 
considering different activities for the two types of monomers, as is usual
for equilibrium polymerization and living polymers. If the mean number of
monomers in the chains is fixed, the entropy as a function of the density of
monomers displays a maximum and is limited from below by the entropy of
monodisperse chains. The increase of entropy due to the polydispersivity is a
linear function of the density of monomers. We also calculate the distribution
of chain sizes, obtaining an exponential relation.
\end{abstract}

\maketitle

\section{Introduction}
\label{intro}
In the microcanonical ensemble, the entropy of a system may be calculated by
evaluating the number of microscopic configurations as a function of the
internal energy and other extensive variables, such as the volume and the
number of particles. The entropy, expressed in terms of these variables, is a
fundamental equation of the system and thus all its thermodynamic properties
may be obtained from this equation. A particular problem of this kind is the
study of the thermodynamic properties of flexible chains placed on regular
lattices. This problem has a rather long history, which reaches back to the
thirties, where the thermodynamics behavior of diatomic molecules
(dimers) adsorbed on surfaces was considered by Fowler and Rushbrooke
\cite{fr37}. In the simplest version of the model, only excluded volume
interactions are taken into account, and thus each site of the lattice
may be either empty or occupied by a single monomer. In this case, the
internal energy of the system will be constant and the temperature is not
defined. The dimer problem may be generalized by considering chains with more
than two monomers. More precisely, we may define the entropy of a system of
chains with $M$ monomers each ($M$-mers) placed on a lattice with $V$ sites as
\begin{equation}
s_M(\rho)=\frac{1}{V}\lim_{V \to \infty} \ln \Gamma(N_p,M;V),
\end{equation}
where $\Gamma(N_p,M,V)$ is the number of ways to place $N_p$ chains with $M$
monomers in each on the lattice 
and the thermodynamic limit is taken with fixed density of occupied sites
$\rho=N_p M/V$. In the particular case of dimers ($M=2$) and fully covered
($\rho=1$) two dimensional lattices, the entropy was calculated exactly
\cite{k61}, in 
other cases estimates were found through series expansions \cite{ncf89},
closed form approximations \cite{so90} and transfer matrix calculations with
finite size extrapolations \cite{ds03}. 

Here we will study the one-dimensional version of the problem. After
discussing briefly the monodisperse case, where all chains are composed by
precisely $M$ 
monomers, we include the possibility of polydispersivity, allowing for a
distribution of different numbers of monomers in chains. In particular, we
consider an ensemble of polydisperse chains determined by different activities
for endpoint- and internal monomers of the chains, such as is usual in models
of equilibrium polymerization and living polymers \cite{wkp80,dfd99}. Perhaps
the most studied experimental realization of equilibrium polymerization is the
phase transition in liquid sulphur, and a good review of this and other
experimental systems is due to Greer \cite{g98}.

In the following section we will define the problem in more detail and show
how it may be solved in the microcanonical ensemble. Section \ref{gc} presents
the calculation in the grand-canonical ensemble, using a transfer
matrix. Besides obtaining the entropy, we calculate also the distribution of
chain sizes. Final discussions may be found in the conclusion.

\section{Definition and microcanonical solution of the model}
\label{calc}
\begin{figure}[h]
\includegraphics[scale=0.4]{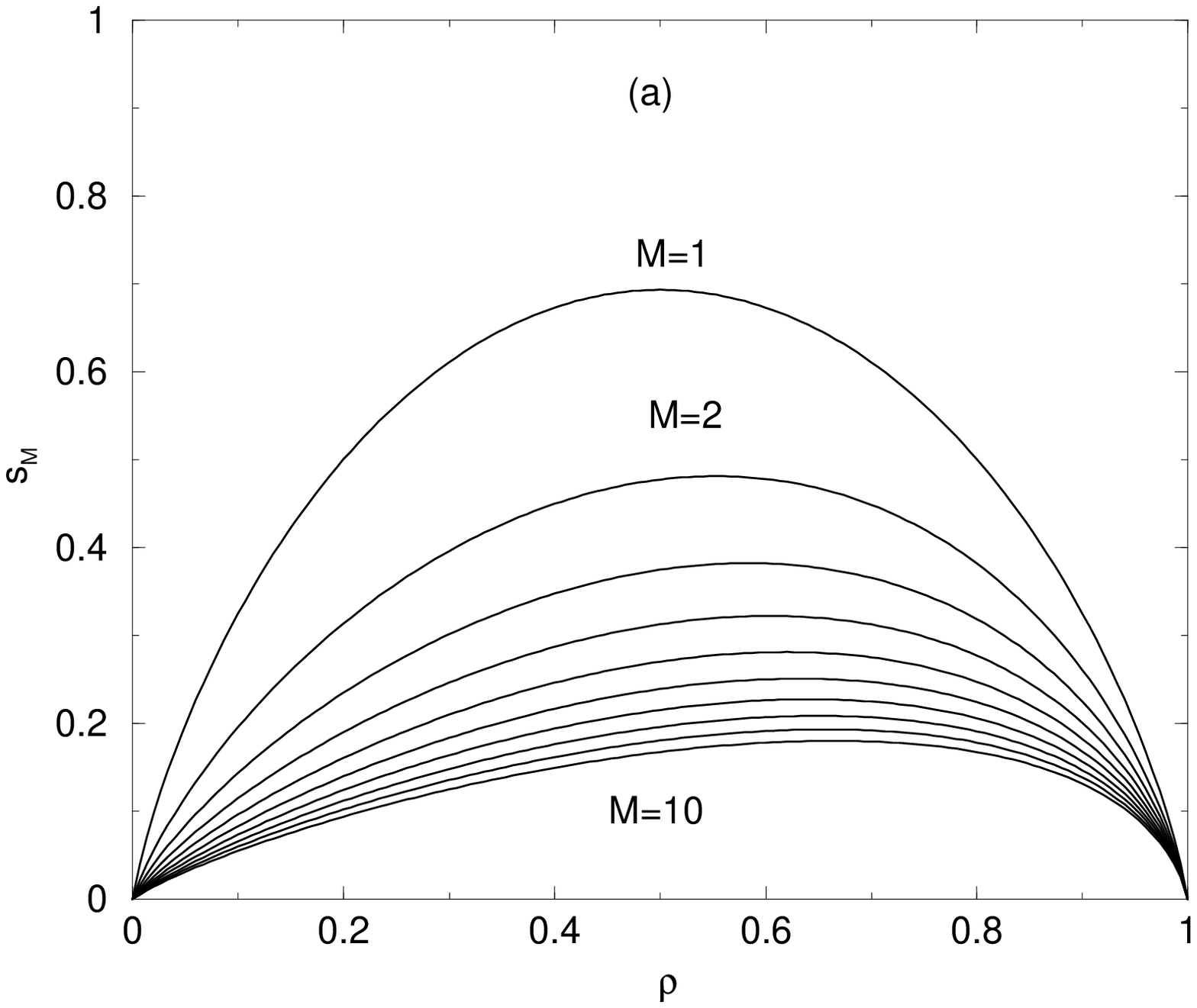}
\includegraphics[scale=0.4]{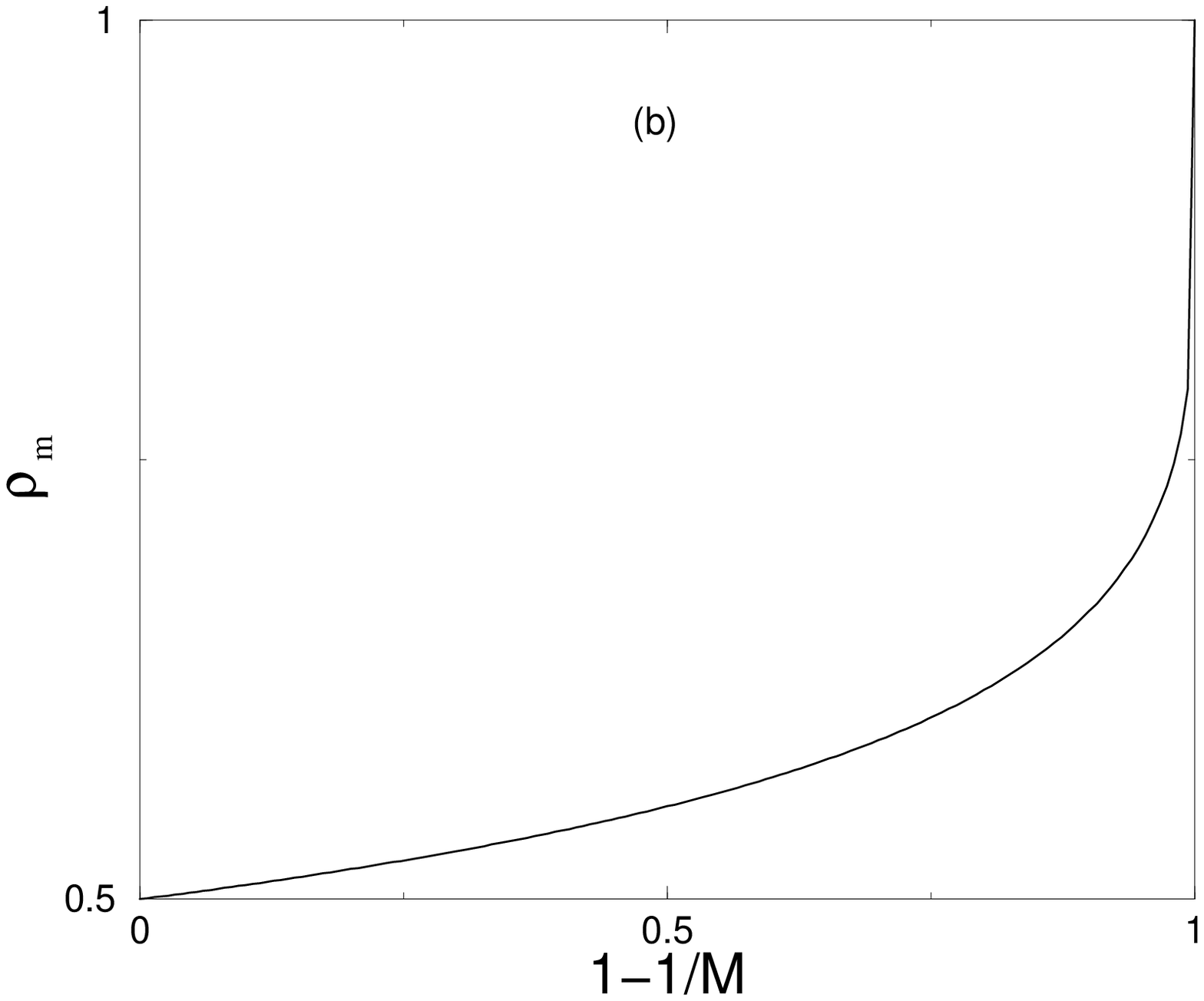}
\caption{a)Entropy as a function of the fraction of occupied sites for
monodisperse chains with molecular weight $M$ between 1 and 10. b)Density of
maximum entropy as a function of $1-1/M$.}
\label{sm}
\end{figure}

Let us start with the formulation and solution of the monodisperse case. In
this case we must calculate the number of ways to place $N_p$ chains with $M$
monomers each on a one-dimensional lattice with $L$ sites. Since in this case
there are $L-M N_p$ empty sites, it is easy to conclude that this number is
equal to
\begin{equation}
\Gamma(N_p,M;L)=\frac{(N_p+L-MN_p)!}{N_p! (L-MN_p)!}.
\label{gamma}
\end{equation}
Taking the thermodynamic limit ($L \to \infty$) with the fraction of occupied
sites $\rho=MN_p/L$ kept fixed and using Stirling's asymptotic form, we get
\begin{eqnarray}
s_M(\rho)&=&\left(\frac{\rho}{M}+1-\rho\right)\ln
\left(\frac{\rho}{M}+1-\rho\right)\nonumber\\
& &-\frac{\rho}{M} \ln 
\frac{\rho}{M} - (1-\rho)\ln(1-\rho).
\label{s1d}
\end{eqnarray}
In Fig. \ref{sm} some curves of the entropy as a function of the density
$\rho$ are shown. The entropy vanishes both at $\rho=0$ and $\rho=1$ and has a
maximum at a density $\rho_m$ which grows monotonically with the molecular
weight $M$, as may be also seen in the figure. In the polymer limit $M \to
\infty$ we find $\rho_m=1$, but the entropy vanishes identically.

\begin{figure}[h]
\includegraphics[scale=0.4]{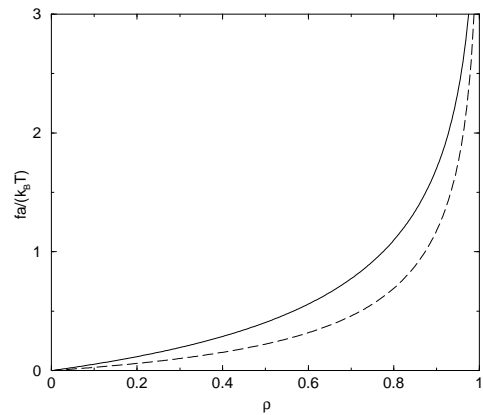}
\caption{Mechanical equation of state for dimers (full line) and tetramers
(dashed line).}
\label{eqst}
\end{figure}

The entropy per site as a function of the density is a fundamental equation of
the system, and all thermodynamic properties of this gas of $M$-mers may be
obtained from it. Being athermal, the temperature is not defined for this gas,
but the intensive entropic variables ($f/T$ and $\mu/T$) may be obtained
taking partial derivatives of $s_M(\rho)$. As an illustration, we will obtain
the mechanical equation of state explicitly. Considering the lattice constant
to be equal to $a$, the physical entropy will be $S=k_B aL s_M(\rho)$, where
$k_B$ is the Boltzmann constant. Since
the system is one-dimensional, the pressure will be replaced by the force $f$
applied to the chain, therefore
\begin{equation}
\frac{f}{T}=\frac{\partial S}{a\partial L}=\frac{k_B}{a}\left(s_M(\rho)-\rho
\frac{d s_M}{d \rho} \right).
\end{equation}
The result is
\begin{equation}
\frac{fa}{k_BT}=\ln\left[1+\frac{\rho}{M(1-\rho)}\right].
\end{equation}
It is interesting to expand the right side of this equation in powers of
$\rho$, and the first terms are
\begin{eqnarray}
\frac{fa}{k_BT} &=& \frac{\rho}{M}+\frac{2M-1}{2}\left(\frac{\rho}{M}\right)^2
\nonumber \\
& & +\frac{1-3M+3M^2}{3}\left(\frac{\rho}{M}\right)^3+\cdots
\end{eqnarray}
The first term corresponds to the ideal gas result, being equal to the number
of $M$-mers divided by $L$. The higher order terms are all positive. In Fig. 
\ref{eqst} the equations of state for dimers and tetramers are plotted. The
linear behavior at low densities and the divergence as $\rho \to 1$ are
apparent. 

In equilibrium polymerization or living polymers, the chains grow through a
process in two steps: first a chain is started and then it grows incorporating
an additional monomer each time. In the case of liquid sulphur, the first step
corresponds to the opening of a small ring if atoms (mainly $S_8$), followed by
the opening and addition of other rings to the linear chain. This process
may be parametrized by two activities or chemical equilibrium constants. The
first one, $z^\prime=\exp(\mu^\prime/k_BT)$, is associated to the initial
process and will be 
considered the statistical weight of chain endpoints. The second activity,
$z=\exp(\mu/k_BT)$, relates to the addition of a monomer to an existing
chain, and will be 
the statistical weight of internal monomers of the chains. $\mu$ and
$\mu^\prime$ are the chemical potentials of internal and endpoint monomers,
respectively
\cite{wkp80}. Usually the propagating reaction which implies growing of the
chains is faster than the opposite depropagating reaction, and this allows for
the possibility of synthesizing chains with a very narrow distribution of
molecular weights by artificially terminating the polymerization process,
through the addition of an appropriate monomer
\cite{g98}. If the equilibrium situation is reached, however, a polydisperse
set of chains will be attained, and this is described by the simple model we
consider with two activities.

In the particular case of sulphur, the endpoint monomer activity $z^\prime$ has
a very small value, of the order of $10^{-12}$, and therefore very long chains
are formed in this polymerization process. The equilibrium polymerization model
may be mapped onto the $n$-vector model of magnetism, and the activity
$z^\prime$ will be proportional to the magnetic field of the $n$-vector model
in the limit $n \to 0$. Thus, in this case the polymerization is a
continuous phase transition \cite{wkp80}. For finite values of $z^\prime$,
where finite chains are dominant, no phase transition occurs, as we expect for
a lattice gas of chains with excluded volume interactions only.

Now let us return to the calculation of the entropy in the one-dimensional
model of equilibrium polymerization. The densities of endpoint monomers $x$ and
internal monomers $y$ are conjugated to the activities $z^\prime$ and $z$,
respectively. All configurations with the same numbers of chains $N_p=Lx/2$
and internal monomers $N_i=Ly$ have the same statistical weight. We may then
calculate the number of these configurations in a combinatorial way. Besides
the multiplicity in Eq. (\ref{gamma}), an additional factor corresponding
to the number of ways to place the internal monomers in the $N_p$ chains is
now present. The result is
\begin{eqnarray}
\Gamma(N_p,N_i;L) &=& \frac{(L-N_p-N_i)!}{(L-2N_p-N_i)!\;N_p!} \times 
\nonumber \\
 & & \frac{(N_p+N_i-1)!}{(N_p-1)!\;N_i!}.
\label{gammap}
\end{eqnarray}
In the thermodynamic limit, we find the following expression for the entropy
per site
\begin{eqnarray}
s(x,y) &=& (1-x/2-y)\ln(1-x/2-y)-  \nonumber \\
& &(1-x-y)\ln(1-x-y)-x\ln(x/2)- \nonumber \\
& &y\ln y+(x/2+y)\ln(x/2+y).
\label{spd}
\end{eqnarray}

The mean number of monomers per chain will be $\bar{M}=2+2y/x$, and if this
number is fixed, we may express the entropy in the polydisperse case as a
function of the density of monomers $\rho=x+y$. The result is
\begin{eqnarray}
s_{\bar{M}}(\rho) &=& \frac{\bar{M}(1-\rho)+\rho}{\bar{M}}\ln
\left[\frac{\bar{M}(1-\rho)+\rho}{\bar{M}}\right]-
\nonumber \\
& &(1-\rho)\ln(1-\rho)
-\frac{2\rho}{\bar{M}}\ln \left[\frac{2\rho}{\bar{M}} \right]-
\frac{\rho(\bar{M}-2)}{\bar{M}} \times  \nonumber \\
& &\ln \left[ \frac{\rho(\bar{M}-2)}{\bar{M}}\right]+
\frac{\rho(\bar{M}-1)}{\bar{M}} \ln \left[ \frac{\rho(\bar{M}-1)}{\bar{M}}
\right].
\end{eqnarray}
As expected, the entropy for the polydisperse case is never smaller than the
one for the monodisperse case when $\bar{M}=M$. The difference between these
entropies is 
\begin{eqnarray}
s_{\bar{M}}(\rho)-s_M(\rho) &=& \frac{\rho}{M}[(M-1)\ln(M-1)\nonumber \\
& &-(M-2)\ln(M-2)].
\end{eqnarray}
The origin of the additional entropy in the polydisperse case is the second
factor in expression \ref{gammap}.

Another point in this problem which may be addressed is the distribution of
sizes of the chains. Again it is possible to answer this in a combinatorial
way, if we consider the probability that a particular chain has $k=M-2$
internal 
monomers. This probability corresponds to the ratio between the number of
internal monomer configurations when one of the $N_p$ chains has exactly $k$
internal monomers and the total number of configurations, which corresponds to
the second factor in Eq. (\ref{gammap}). In the thermodynamic limit, this
leads to the result
\[
r_k=\frac{x}{x+2y}\left[ \frac{2y}{x+2y} \right]^k,
\]
with $k=0,1,2,\ldots$. This result may be rewritten as
\begin{equation}
r_M=\frac{1}{\bar{M}-1}\left[\frac{\bar{M}-2}{\bar{M}-1}\right]^{M-2},
\label{dist}
\end{equation}
where $M=2,3,4,\ldots$ and $r_M$ is the fraction of chains with $M$ monomers,
which is a function of the mean number of monomers in chains $\bar{M}$.
It is worth to notice that the Flory-Huggins
approximation to a lattice model for equilibrium polymerization also results
in an exponential distribution of the chain sizes \cite{dfd99}. Both results
are in fact equivalent, if one takes into account that Dudovicz {\it et al.}
considered isolated sites to be chains with one monomer, while here we called
them empty sites.

\section{Transfer matrix solution in the grand-canonical ensemble}
\label{gc}
\begin{figure}[h]
\includegraphics[scale=0.6]{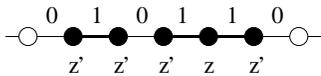}
\caption{Section of the lattice with a particular configuration of chains. The
values of the bond variable $\eta$ are indicated above the line and below the
line the activity of each monomer is given.}
\label{lt}
\end{figure}

The grand-canonical solution of the monodisperse case may be found in
reference \cite{ds03}, so the polydisperse case will be discussed here. The
equilibrium polymerization model in one dimension was discussed already in
reference \cite{pw83}, with emphasis on the limit of infinite chains, where a
phase transition occurs. Here we will use a similar transfer matrix technique
to obtain the entropy and the distribution of chain sizes. To define a
transfer matrix, we may use lattice gas variables $\eta_i=0,1$ associated to
the bonds of the lattice, such that $\eta_i=1$ if the bond $i$ is inside a
chain and $\eta_i=0$ otherwise. This definition is illustrated in Fig.
\ref{lt}, where a section of the lattice is depicted.

The transfer matrix for this case is given by
\begin{equation}
T=\left(
\begin{array}{cc}
1 & z^\prime \\
z^\prime & z
\end{array}
\right).
\end{equation}
The largest eigenvalue of this matrix is
\begin{equation}
\lambda_1=\frac{1+z+\sqrt{(1-z)^2+4(z^\prime)^2}}{2}.
\end{equation}
This eigenvalue may become degenerated when $z^\prime$ vanishes, originating
the polymerization transition studied in reference \cite{pw83}. For our
present interest, we will consider finite values of $z^\prime$, so that no
degeneracy occurs. In the thermodynamic limit, the grand-canonical potential
will be given by
\begin{equation}
\Phi=-k_BTV\ln \lambda_1,
\end{equation}
and since the entropy $S$ is the partial derivative of $\Phi$ with respect to
the temperature, and remembering that $z=\exp(\mu/k_BT)$ and
$z^\prime=\exp(\mu^\prime/k_BT)$, we obtain the following expression for the
adimensional entropy per site 
\begin{equation}
s=\frac{S}{k_BV}=\ln \lambda_1-\frac{z}{\lambda_1}\frac{\partial
\lambda_1}{\partial z} -\frac{z^\prime}{\lambda_1}\frac{\partial
\lambda_1}{\partial z^\prime}.
\end{equation}
Now we may recognize that 
\begin{equation}
y=\frac{z}{\lambda_1}\frac{\partial \lambda_1}{\partial z}
\end{equation}
is the density of internal monomers and
\begin{equation}
x=\frac{z^\prime}{\lambda_1}\frac{\partial \lambda_1}{\partial z^\prime}
\end{equation}
is the density of endpoint monomers in chains. Some algebra will lead us to
expression \ref{spd} for the entropy in the polydisperse case. In the limit
$y \to 0$ we have $x=\rho$ and the entropy of dimers $s_{M=2}(\rho)$ is
recovered.

\begin{figure}[h]
\includegraphics[scale=0.6]{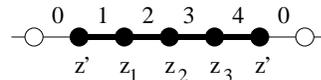}
\caption{Section of the lattice with a particular configuration of chains. The
bond variables $\eta$ and monomer activities are shown, defined in a way to
allow for the calculation of the size distribution of chains.}
\label{lattd}
\end{figure}

To obtain the distribution of internal monomers among the chains, we may
define an activities for the internal monomers which are dependent on the
localization of the monomer in the chain. Thus, the leftmost internal monomer
will have an activity $z_1$, his neighbor to the right has an activity $z_2$
and so on. The bonds of the lattice will also be numbered accordingly. A bond
which is not inside a chain is represented by $\eta=0$, as before, but the
$i$'th internal bond, counted from the left to the right, is associated to
$\eta=i$. These definitions are illustrated in Fig. \ref{lattd}.

The transfer matrix for the model now has an unlimited size, but a very simple
structure
\begin{equation}
T=\left(
\begin{array}{ccccc}
1&z^\prime&0&0&\cdots \\
z^\prime&0&z_1&0&\cdots\\
z^\prime&0&0&z_2&\cdots\\
\vdots&\vdots&\vdots&\vdots&\vdots\\
\end{array}
\right).
\end{equation}
The secular equation may be found developing the determinant $|T-\lambda I|$
recursively by the first line. It is more convenient to express the equation
in terms of $\omega=1/\lambda$, and after some algebra, we get
\begin{equation}
-1+\omega+(z^\prime\omega)^2\left(1+\sum_{i=1}^\infty \prod_{j=1}^i
z_j\omega^i\right)=0. 
\end{equation}
Now the density of monomers in the $k$'th position in chains will be
\begin{equation}
\rho_k=\frac{z_k}{\lambda}\frac{\partial \lambda}{\partial z_k}=
-\frac{z_k}{\omega}\frac{\partial \omega}{\partial z_k}.
\end{equation}
Deriving the secular equation with respect to $z_k$ and then making all $z_k$
equal to $z$, we get
\begin{equation}
\rho_k=\frac{(z^\prime)^2 \omega(1-\omega z)}{(1-\omega z)^2+(z^\prime)^2
\omega(2-\omega z)}(\omega z)^k.
\end{equation}
This result may be written as a function of the monomer densities $x$ and $y$
\begin{equation}
\rho_k=\frac{x}{2}\left(\frac{2y}{x+2y}\right)^k.
\end{equation}
From this expression, the fraction $r_k$ of chains with $k$ internal monomers
expressed in Eq. (\ref{dist}) may easily be found.

\section{Conclusion}
\label{conc}
 
Although it was rather simple to find the entropy and distribution of chain
sizes in the equilibrium polymerization model using combinatorial arguments in
the microcanonical ensemble, it is interesting to perform these calculations
also in the grand-canonical ensemble, since the combinatorial calculations are
difficult to generalize to more complex situations, such as the problem
defined on strips of finite widths (ladders). For the monodisperse case, such
calculations have lead to rather precise estimates in the two-dimensional
limit \cite{ds03}. We are presently performing similar calculations for
directed and self-avoiding chains on strips, which hopefully will allow us to
estimate the properties of these models in the two-dimensional limit.

\begin{acknowledgments}
This work was supported by the brazilian agencies CNPq and FAPERJ. JFS
acknowledges funding by project Pronex-CNPq-FAPERJ/171.168-2003.
\end{acknowledgments}

\end{document}